\documentclass[12pt]{iopart}

\usepackage{graphicx}
\usepackage{epsfig}
\usepackage{amssymb}
\usepackage{bm}

\begin{document}

\title[Fitness collapse in spatial population genetics]{Critical Fitness Collapse in Three-Dimensional Spatial Population Genetics}

\author{Maxim O. Lavrentovich}

\address{Department of Physics \& Astronomy, University of Pennsylvania, Philadelphia, PA 19104 USA}
\ead{lavrentm@gmail.com}
\vspace{10pt}
\begin{indented}
\item[]April 2015
\end{indented}

\begin{abstract}

  If deleterious mutations near a fitness maximum in a spatially distributed population  are sufficiently frequent or detrimental, the population can undergo a fitness collapse, similarly to the Muller's ratchet effect in well-mixed populations.  Recent studies of one-dimensional habitats (e.g., the frontier of a two-dimensional range expansion) have shown that the onset of the fitness collapse is described by a directed percolation phase transition with its associated critical exponents.  We consider population fitness collapse in three-dimensional range expansions with both inflating and fixed-size frontiers (applicable to, e.g., expanding and treadmilling spherical tumors, respectively).   We find that the onset of fitness collapse in these two cases  obeys different scaling laws, and that competition between species at the frontier leads to a deviation from directed percolation scaling.  As in two-dimensional range expansions, inflating frontiers modify the critical behavior by causally disconnecting well-separated portions of the population. 

\end{abstract}

%
\vspace{2pc}

%
%
%

\section{\label{SIntro}Introduction}

    Typical laboratory evolution experiments, especially in microbial populations \cite{lenski}, are conducted in a well-mixed environment.   Theoretical treatments of evolution also often focus primarily on the well-mixed case \cite{Ewens}.    However, in nature, populations will usually have some spatial variation.  The spatial distribution of genotypes is particularly important in range expansions, in which populations invade new territory \cite{KorolevRMP}.  For example, cancer cells may form solid tumors which invade healthy tissue \cite{cancermodel}, microbial colonies may spread over a surface such as a Petri dish \cite{KorolevBac,KorolevMueller}, and animals like butterflies and birds  settle new territory over time \cite{animalRE}.  A key  impact of spatial variations on evolution is an enhanced noise  (called genetic drift) due to number fluctuations in local regions with small effective population sizes at the frontier.   This genetic drift can lead to fluctuation-driven phase transitions, characterized by the extinction of a particular strain within a population.  This paper focuses on fitness loss in populations due to such phase transitions in three-dimensional range expansions.

Previous work  has focused on evolutionary dynamics in populations with one-dimensional frontiers  \cite{spatialratchet,MKNPRE,KorolevRMP,nelsonhallatschek}.  The spatial dynamics and geometry of the population strongly influences its evolutionary dynamics, creating dramatic changes relative to a well-mixed population.  For example, Krug and Otwinowski \cite{spatialratchet} have found that spatial fluctuations can  lead to a fitness collapse that is reminiscent  of Muller's ratchet in well-mixed populations \cite{ratchet1,ratchet2}, in which the average fitness of the population  declines as its fittest members go extinct via the combined effects of genetic drift and deleterious mutations. Effectively one-dimensional habitats may be realized, for example, at the frontier of a two-dimensional range expansion, such as a microbial colony on a Petri dish \cite{KorolevRMP, MKNPRE}.     In such expansions, growth is often confined to a thin layer at the population frontier.

\begin{figure}[!ht]
\centering
\includegraphics[width=3.5in]{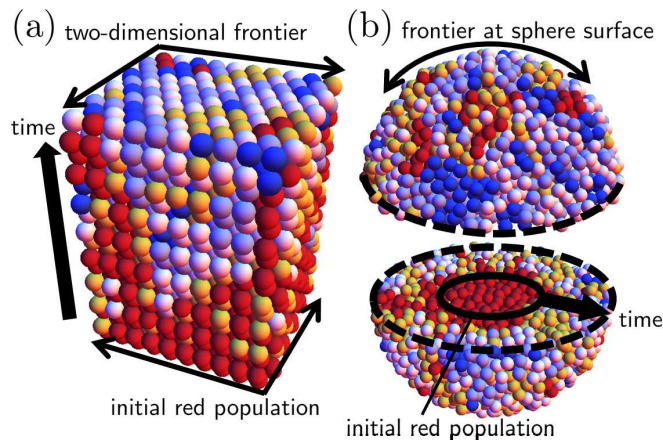}
\caption{ \label{fig:simintro}(a) A  small simulated three-dimensional range expansion (with periodic boundary conditions perpendicular to the time-like growth direction) and an initial population of all red cells, representing a maximally fit lineage.  As the cells divide and populate the subsequent generations (the two-dimensional sheets of cells in a triangular lattice along the vertical direction) they acquire deleterious mutations (each decreasing the growth rate by 10\%) with a 20\% probability per division. The cells gradually turn more blue as they acquire more mutations. (b)  A simulated spherical range expansion with the same deleterious mutations, cut in half to reveal the initial red population (about six cells in diameter) at the center.  The dashed black lines indicate the edges of the two hemispheres. The generations are concentric spherical shells about a single cell wide. The population frontier inflates in such an expansion:  The initial ball of red cells  grows to a cluster with about a 20 cell diameter.   Section~\ref{sec:model} discusses the simulations in more detail.    }
\end{figure}

 In this paper, we study the evolutionary dynamics at the edge of three-dimensional range expansions such as tumors, spherical microbial colonies grown in soft gels on liquid media, etc.   We will compare expansions of populations experiencing deleterious mutations  at non-inflating fronts, as in figure~\ref{fig:simintro}(a), with those living on curved, inflating frontiers, as in figure~\ref{fig:simintro}(b).  In both cases, we focus on expansions with a single-cell-wide frontier of active growth.  This limit maximizes the effects of genetic drift and is a realistic model of, e.g., tumor populations which may have thin actively growing regions \cite{hochberg}.       To make contact with non-equilibrium statistical physics literature \cite{NEQPTBook,Hinrichsen}, it is convenient to refer to such expansions as $d=2+1$-dimensional expansions (or $d=1+1$ for two-dimensional expansions with thin frontiers), where the  $2$ is the effective frontier dimension, and the ${}+1$ refers to the ``time-like'' direction in which the expansion spreads, as shown in figure~\ref{fig:simintro}.

  Deleterious mutations often accumulate in genetically unstable populations, such as tumors with a high mutational load  \cite{KorolevCancer} or in viral populations, with their imperfect proofreading machinery \cite{viralmut}.    Understanding the nature of this accumulation is important, for example, for designing effective cancer therapies   \cite{KorolevCancer}. Since the target space for deleterious mutations is typically much larger than for beneficial ones, there may be time scales over which populations almost exclusively experience irreversible, deleterious mutations. Hence, we will focus on the accumulation of such effectively one-way mutations and the onset of fitness collapse that can result.   As discussed below, the collapse can be associated with a phase transition, slightly smeared by finite size effects. Much like at a second-order equilibrium phase transformation, we expect that universal behavior appears near this transition, with results insensitive to the microscopic details of our models \cite{NEQPTBook}.  Indeed, many diverse systems can share a single \textit{universality class} and identical scaling characteristics near the transition.  For example, Otwinowski and Krug \cite{spatialratchet} showed that the onset of fitness collapse at the one-dimensional frontier of a flat (i.e., non-inflating) two-dimensional range expansion  is governed by the directed percolation (DP) universality class, which describes a wide variety of systems, ranging from growing interfaces to turbulent liquid crystals \cite{Hinrichsen, NEQPTBook}. 
We describe here the analogous onset in three-dimensional range expansions and find important differences, including a different phase diagram shape and deviations from conventional DP scaling.  We also consider the effects of an inflating frontier [see figure~\ref{fig:simintro}(b)], which leads to a cutoff in the critical scaling by mitigating some of the effects of genetic drift.    

We model range expansions in which we track  $M$ different types of cells, where $M$ may be finite or infinite.  Each type will represent a cell with $m$ accumulated deleterious mutations, where $m=0,1,\ldots,M-1$. We label the species with different colors, as shown in figure~\ref{fig:simintro}.   The red cells are the fittest cells with $m=0$.  The cells turn more blue as they acquire more mutations.   For flat frontiers, as in figure~\ref{fig:simintro}(a), we will study how increasing $M$ influences the onset of fitness collapse.  To study the effects of inflating frontiers, as in figure~\ref{fig:simintro}(b), we will focus on the $M=2$ case.  An experimental realization of our models  is within reach:  Current advances in fluorescence detection have greatly expanded our ability to track multiple lineages of cells in spatially distributed populations using confocal fluorescence microscopy.  In fact, tracking $M \approx 100$ lineages using as many different fluorescent reporters may be feasible \cite{brainbow1,brainbow2,brainbow3}.  As such tools become more widely used, it is of interest to map out the theoretical space of possibilities for the evolutionary dynamics.

This paper is organized as follows: In section~\ref{sec:model}, we introduce our range expansion lattice model and a coarse-grained description using  stochastic partial differential equations. Section~\ref{sec:meanfield} presents a mean-field analysis of the coarse-grained model which recapitulates some known results for well-mixed populations (such as the quasispecies model of Eigen and collaborators \cite{eigen}) and illustrates various possibilities for the critical behavior.  We present results for the onset of a fitness collapse in three-dimensional range expansions for large values of $M$ with flat fronts [as in figure~\ref{fig:simintro}(a)] and compare to two-dimensional expansions in section~\ref{sec:dp}.  We focus on changes due to inflating frontiers for $M=2$ in section~\ref{sec:inflation} and conclude in section~\ref{sec:conclusions}.

\section{\label{sec:model} The Model}

\begin{figure}[!ht]
\centering
\includegraphics[width=2.5in]{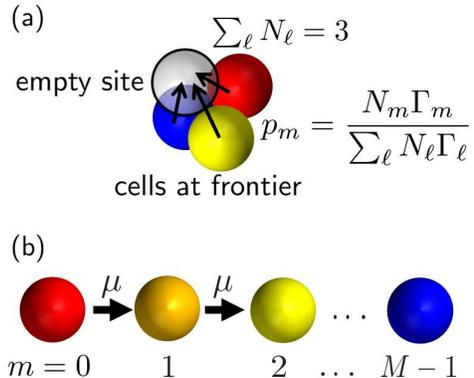}
\caption{ \label{fig:model}(a) A single update step in our lattice model of planar range expansions.  Cells at the frontier compete to divide into the (uncolored) adjacent empty site.  The site gets filled with a cell of fitness class $m$ with probability $p_m$, determined by $\Gamma_m = (1-s)^m$ and $N_m$,  the number of competing cells with class $m$.  In this case, three total cells are competing, so $\sum_{\ell} N_{\ell}=3$. (b)  After this competition, the daughter cell can acquire an additional one-way deleterious mutation with probability $\mu$, moving it from class $m$ to $m+1$.  The most deleterious class is $m=M-1$. }
\end{figure}

The expansions in figure~\ref{fig:simintro} were simulated using a variation of the models presented in \cite{MKNPRE, MLSphere}.   In these expansions, deleterious mutations accumulate at a constant rate $\mu $ per cell division.  Each mutation has a multiplicative fitness cost, and the growth rate $\Gamma_m$ of a cell with $m$ acquired mutations is thus
\begin{equation}
\Gamma_m = (1-s)^m, \label{eq:growthrates}
\end{equation}
where $s$ $(0<s<1)$ is the strength of the deleterious effect, and the ``fitness class'' $m$ ranges from $0$ to $M-1$, so that  $M$ is the total number of cell types.  Note that $\Gamma_m$ is always positive, and that the growth rate when no deleterious mutations are present has been rescaled to unity.    In the small $s$ limit, our model amounts to a spatial generalization of a well-mixed population genetics problem studied in \cite{generallangevin}.  To simulate planar fronts in a three-dimensional expansion, we stack two-dimensional ``sheets'' of cells in a triangular lattice (with periodic boundary conditions) to form  a three-dimensional hexagonal close packed (hcp) lattice. Each two-dimensional sheet of cells represents a single generation [see figure~\ref{fig:simintro}(a)].  Each approximately square sheet will have $N$ cells arrayed along the two dimensions of the sheet.  At the frontier, the structure of the hcp lattice ensures that each empty site is adjacent to three frontier cells, all of which compete to divide into that site, as shown in figure~\ref{fig:model}(a).  These three frontier cells divide with a probability proportional to their growth rate.   In particular, a cell of type $m$ is placed in the empty site with a normalized probability $p_m \propto N_m \Gamma_m$, where $N_m$ is the number of competing frontier cells with class $m$.    Upon invoking a normalization condition, we have $p_m = N_m \Gamma_m/\sum_{\ell} N_{\ell} \Gamma_{\ell}$. After a new daughter cell is placed in the empty spot, it can acquire an additional deleterious mutation moving it down a fitness ladder with some probability $\mu$ as shown in figure~\ref{fig:model}(b).

 All empty sites in a two-dimensional sheet are filled before moving on to the next generation, ensuring  flat population fronts for the three-dimensional range expansions.        As in \cite{KorolevRMP}, a \textit{single} triangular lattice sheet may be used to simulate a \textit{two}-dimensional  expansion.  In this case, rows of the triangular lattice represent the successive population frontiers and two cells compete for each empty site.     Such uniform fronts are reasonable approximations when the population has an effective surface tension, such as at the periphery of a yeast colony \cite{brenneryeast}, and when differential growth rates between species at the frontier are small (i.e., $s \ll 1$ in our model).  However, when present, undulations of the population frontier can strongly modify the evolutionary dynamics.  For example, in two-dimensional expansions, deviations from directed percolation scaling were found at the onset of fitness collapse \cite{frey}. Although we do not study them here, undulations could cause deviations in the critical exponents associated with  three-dimensional expansions, as well (see, e.g., \cite{MLDRNPRL}).

 Special techniques are required for circular or spherical expansions to avoid lattice artifacts \cite{MKNPRE}.  For spherical range expansions, we pick an empty site closest to the center of the growing population that neighbors at least one cell along the population frontier.  This update rule creates a uniform front and mimics a  surface tension in the spherical cell cluster \cite{MLSphere}.   The site positions for a disordered lattice are generated in advance using the Bennett algorithm, originally used to model disordered metallic glasses \cite{Bennett} (see \cite{MLSphere} for more details).   Initial conditions in such expansions are specified by assigning particular colors to all cells a distance of $R_0$ or less from the population center.  In our simulations, $R_0$ is measured in cell diameters $a$.  After the expansion grows out to some maximum radius, we break up the resulting ball of cells into concentric spherical shells that are a single cell thick.  These shells then define a temporal succession of population frontiers.  Thus, in both  types of expansions illustrated in figure~\ref{fig:simintro}, the update rules are designed so that the population expands outward by one cell diameter $a$ per generation.  We measure time in generations and choose length units that fix the population front speed to be  $v = a/\tau_g = 1$, where $\tau_g=1$ is a generation time.

We will be interested in the deleterious one-way mutation  regime  $(s,\mu>0)$ where we find a fitness collapse transition.    We will not consider beneficial mutations $(s < 0)$, which have their own interesting phenomenology \cite{spatialratchet}.   Our  mutations are irreversible, so that when the fittest class $m=0$ dies out, it cannot be regenerated.  The system is then in   an absorbing subspace of states from which it cannot escape.   Note that with this model  we can have irreversible extinctions at all class levels $m$ as long as all cells in the fitter classes $n < m$ have died out as well.    When $M=2$, our lattice model for range expansions with planar fronts reduces to a $d=2+1$-dimensional version of the Domany-Kinzel stochastic cellular automaton \cite{domany}.   In this case, there is a single absorbing state in which the entire population is in the less fit $m=1$ class.    Transitions into a single absorbing state, governed by the directed percolation (DP) universality class, are the subject of a wealth of literature in nonequilibrium statistical dynamics \cite{NEQPTBook}.  Hence, as in the original $d=1+1$ model (see \cite{MKNPRE}), we expect that when $\mu,s>0$, there will be a line of DP phase transitions at which we find the onset of fitness collapse \cite{MKNPRE, NEQPTBook, Hinrichsen}.   We will use the $M=2$ case to guide our thinking about the more complicated and biologically relevant limit $M \gg 1$.

To find an analytically tractable description of the evolutionary dynamics, it is useful to move to a coarse-grained model.  Let $f_m \equiv f_m(\mathbf{x},t)$ be the fraction of $m$-class cells  in a small, approximately planar patch  surrounding a point $\mathbf{x}$ on the population frontier at time $t$ \cite{MKNPRE}.       The cells in each local patch are treated as if they are in a well-mixed population.  The cells are allowed to hop to an adjacent patch via a local cell rearrangement (occurring, for example, via outward pushing due to cell division or slight cell motility).    The evolutionary dynamics in each well-mixed patch is modelled by a stochastic differential equation equation, developed by Good and Desai \cite{generallangevin}, that assumes the growth rates in equation~(\ref{eq:growthrates}) in the small $s$ limit, a constant population density, irreversible mutations between classes $m$ and $m+1$ with rate $\mu$, and the genetic drift associated with the finite population size in each patch.  Then, for a non-inflating, flat population frontier, the fractions $f_m $ of $m$-class cells obey the spatial generalization of this equation: 
\begin{eqnarray}
\partial_t f_m & = D \nabla^2 f_m+ s(\langle m \rangle-m) f_m - \mu f_m (1-\delta_{m,M-1}) + \mu f_{m-1}  \nonumber \\ &   \quad {} +\sum_{\ell=0}^{M-1} (\delta_{m,\ell}-f_m) \sqrt{2 \tau_g^{-1}\Delta f_{\ell}} \, \xi_{\ell}(\mathbf{x},t), \label{eq:mclang}
\end{eqnarray}
where  $\delta_{i,j}$ is a Kronecker delta function  ($\delta_{i,j}=1$ if $i=j$ and 0 otherwise), $\langle m \rangle = \sum_{\ell=0}^{M-1} \ell f_{\ell}$ is the local average fitness class, and $\tau_g$ is the generation time. The strength of the genetic drift, $\Delta = N^{-1}$, is  set by the population size $N$ of each local patch.  For each class, the  It\^o noise terms $\xi_{\ell}(\mathbf{x},t)$ \cite{gardiner} have vanishing mean and variances given by $\langle \xi_{\ell}(\mathbf{x},t) \xi_{m}(\mathbf{x}',t') \rangle = \delta_{\ell m} \delta(\mathbf{x}-\mathbf{x}') \delta(t-t')$. The diffusion constant   $D \sim a^2/\tau_g$  describes random cell rearrangement at the frontier, which in our lattice simulations is expected to move cells by approximately a cell diameter $a$ per generation.  We impose the boundary conditions $f_{-1}=0$ and $\sum_{\ell=0}^{M-1} f_{\ell}=1$.       Our model is essentially a stepping-stone model  of the range expansion \cite{kimurapaper, KorolevRMP},  and can be derived more rigorously from a modified, continuous-time version of the lattice model \cite{MKNPRE}.  Note that our lattice simulations are in the limit where each patch population size is $N=1$, which maximizes the effects of genetic drift.

 We will examine equation~(\ref{eq:mclang}) near the fitness collapse transition, where the fractions $f_m$ at the fit edge of the fitness distribution (i.e., the first few values of $m$)  will be small  and close to their absorbing state values $f_{m}=0$.  When the fractions $f_m$ are small,  equation~(\ref{eq:mclang}) reduces to a set of (nonlinear) coupled DP Langevin equations for the first few values of $m$, i.e., directed percolation with ``colors,'' studied extensively in \cite{RougheningCoupledDP, JanssenCoupledDP,TauberCoupledDP}.   To see this, we set $f_{M-1}=1-\sum_{\ell=0}^{M-2}f_{\ell}$ and  keep only the leading order noise terms in equation~(\ref{eq:mclang}).  Then,  the Langevin equations for $0 \leq m \ll M-1$ reduce to: 
\begin{equation}
\partial_t f_{m}\approx D \nabla^2 f_m - \left[ r_m+ \sum_{\ell=0}^{M-2}T_{m \ell}f_{\ell}   \right]f_m+ \mu f_{m-1}+ \sqrt{2 \tau_g^{-1}\Delta f_{m}} \,   \xi_{m}(\mathbf{x},t),\label{eq:UCDPLang}
\end{equation}
where $r_m =~\mu -s(M-1-m)$ and $T_{m \ell}=s(M-1-\ell)$.     One can show that the higher order noise terms we dropped are irrelevant in the renormalization group sense. Equation~(\ref{eq:UCDPLang}) is now a hierarchy of DP  Langevin equations, coupled to each other both linearly, via the mutation term $\mu f_{m-1}$, and bilinearly, via the off-diagonal matrix elements $T_{m \ell} f_m f_{\ell}$ (with $m \neq \ell$)  \cite{RougheningCoupledDP, JanssenCoupledDP,TauberCoupledDP}.     When $M=2$, equation~(\ref{eq:UCDPLang}) reduces to the DP Langevin equation \cite{NEQPTBook} for the $m=0$ class, studied in the context of spatial population genetics in \cite{MKNPRE}.  Hence, for the special case $M=2$, we expect to find DP critical behavior in our model.  We will confirm this result in the next section.    It is possible to generalize equation~(\ref{eq:UCDPLang}) to include the effects of inflating frontiers  \cite{MLSphere, MKNPRE}. Although we study inflation via simulations using a disordered lattice in section~\ref{sec:inflation}, an analytic treatment of the inflationary generalization of equation~(\ref{eq:UCDPLang}) is  beyond the scope of this paper.

  If we allow $r_m$ and $T_{m \ell} $ in equation~(\ref{eq:UCDPLang})  to be arbitrary (instead of given by the expressions just below equation~(\ref{eq:UCDPLang})), this set of equations exhibits a  rich phase structure.  In particular, the parameters $\{r_m\}$ can be tuned so that any, some, or all of the $m$-species  go extinct.  The multicritical point at which all of the species go extinct simultaneously (for $M \rightarrow \infty$) is particularly interesting as it describes the critical dynamics of some interface growth models at a roughening transition \cite{RougheningCoupledDP,spatialratchet}.    The field  $f_m(\mathbf{x},t)$ for the interface models represents the fraction of interface plateaus of height $m$ at some point $\mathbf{x}$ and at time $t$.  A simple Langevin equation exhibiting such a multicritical point was proposed in \cite{RougheningCoupledDP}. The equation has the same form as equation~(\ref{eq:UCDPLang}), but with constant coefficients  $r_m = r$ and a constant, diagonal matrix $T_{m \ell} = T \delta_{m \ell}$.      This equation describing the interface growth models falls into the unidirectionally coupled directed percolation (UCDP) class near the multicritical point \cite{TauberCoupledDP,TauberCoupledDP2}.  We shall see that this UCDP universality class will be relevant for our range expansion evolutionary dynamics, as well.     The presence of the bilinear couplings $T_{m \ell}$ with $m \neq \ell$, however, can lead to deviations from unidirectionally coupled directed percolation scaling \cite{AsymmCoupledDP}. We will now describe a mean-field analysis of equation~(\ref{eq:UCDPLang}) to better understand the phase structure of our model and to point out various possibilities for the critical behavior.

\section{Mean-Field Analysis \label{sec:meanfield}}

\begin{figure}[!ht]
\centering
\includegraphics[width=2.5in]{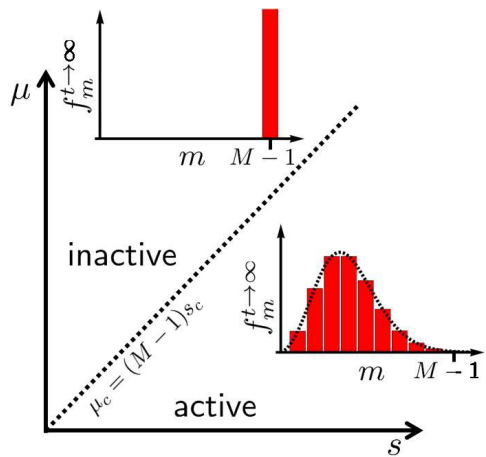}
\caption{ \label{fig:ratchetMF}  An error threshold transition for a well-mixed population.  When $\mu/s < M-1$, we are in an ``active'' phase where the population has a fixed fitness distribution at long times, determined by mutation-selection balance. The distribution is given approximately by equation~(\ref{eq:poissonstat}), i.e. a Poission distribution with mean $\mu/s$ (lower inset).  When $\mu/s>M-1$, the fitness distribution collapses and the entire population is eventually in the least fit class for long times. }
\end{figure}

If we remove the noise (i.e. genetic drift) and diffusion terms in equation~(\ref{eq:UCDPLang}) by setting $D=\Delta=0$, we recover the mean-field limit.  This limit also corresponds to the dynamics of an infinite ($N \rightarrow \infty$), well-mixed population.  In this case, our fitness class fractions $f_m(\mathbf{x},t)$ are independent of $\mathbf{x}$ and evolve deterministically.  Hence, we cannot have noise-driven extinction events.  However, there is still a phase transition for any finite total number of species $M< \infty$.  Since this transformation is not noise-driven, it is not a Muller's ratchet and is more properly called an ``error threshold'' transition \cite{threshold1}, a term originating from the theory of quasi-species \cite{eigen}.    At and above this threshold, the mutation rate is high enough to induce a fitness distribution collapse in which the entire population eventually collapses into the least fit $m=M-1$ class (largest value of $m$), as illustrated in figure~\ref{fig:ratchetMF}.

 We can solve for the mean-field stationary fractions  $f_m^{t \rightarrow \infty} \equiv f_m(\mathbf{x},t \rightarrow \infty)$ using the method of generating functions \cite{functionology}. The asymptotic long time fractions $f_m^{t \rightarrow \infty}$ obey the steady-state, mean field version of equation~(\ref{eq:UCDPLang}):
\begin{equation}
\mu f^{t \rightarrow \infty}_{m-1}-\left[ r_m + \sum_{\ell=0}^{M-2}T_{m \ell} f^{t \rightarrow \infty}_{\ell} \right]f^{t \rightarrow \infty}_m=0, \label{eq:fitrecursion}
\end{equation}
 with  $r_m =~\mu -s(M-1-m)$ and $T_{m \ell}=s(M-1-\ell)$. The generating function packages information about the different fitness classes into a polynomial function of $x$, defined by
\begin{equation}
G(x) \equiv \sum_{\ell=0}^{M-1} f_{\ell}^{t \rightarrow \infty} x^{\ell}.
\end{equation}
Equation~(\ref{eq:fitrecursion}) simplifies to  a differential equation for  $G(x)$ when we multiply both sides by $x^{\ell}$ and sum over all possible $\ell$:
\begin{equation}
\frac{dG}{dx} = \frac{\mu}{s}\, G(x)+ \frac{\mu}{s} \left[ \frac{1}{x}-1\right]f_{M-1}^{t \rightarrow \infty} x^{M-1}, \label{eq:diffeqgenf}
\end{equation}
with the boundary condition $G(1)=1$.  The last term in equation~(\ref{eq:diffeqgenf}), proportional to  the fraction of cells in the last allele class, $f_{M-1}^{t \rightarrow \infty}$,  can be reexpressed in terms of $G(x)$ using the  relation
\begin{equation}
f_m^{t \rightarrow \infty} = \frac{1}{m!} \, \left. \frac{d^m}{dx^m}G(x) \right|_{x=0}. \label{eq:geninverse}
\end{equation}
Upon solving equation~(\ref{eq:diffeqgenf}) for $G(x)$, we use equation~(\ref{eq:geninverse}) to find the steady-state fitness distribution.  Equation~(\ref{eq:diffeqgenf}) is most readily solved when  the fraction $f_{M-1}^{t \rightarrow \infty}$  is small, or if $M \rightarrow \infty$. Then, we can  ignore the   term proportional to $f_{M-1}^{t \rightarrow \infty}$ in equation~(\ref{eq:diffeqgenf}) and find that  $G(x) = e^{(\mu /s)(x-1)}$, which is the generating function for  a  Poisson distribution:
\begin{equation}
f_m^{t \rightarrow \infty} = \frac{1}{m!} \left( \frac{\mu}{s} \right)^m e^{-\mu/s} \qquad \qquad (f_{M-1}^{t \rightarrow \infty} \ll 1 \mbox{ or } M \rightarrow \infty). \label{eq:poissonstat}
\end{equation}
Equation~(\ref{eq:poissonstat}) will be a good approximation to the steady-state distribution as long as $\mu/s \ll M-1$ in a well-mixed population, i.e. far below the dashed transition line in figure~\ref{fig:ratchetMF}.  For spatial range expansions, however, the steady-state distribution will be quite different.  To highlight this difference, we will calculate $f_0^{t \rightarrow \infty}$ (the fraction of the population in the most fit class) for the range expansions in the next section and compare to the well-mixed population result from equation~(\ref{eq:poissonstat}): $f_0^{t \rightarrow \infty} = e^{-\mu/s}$.

 When $M < \infty$, equation~(\ref{eq:fitrecursion}) exhibits a line of phase transitions in the $(s,\mu)$-plane at $r_0 \equiv \mu -s(M-1)=0$, i.e. for $\mu/s = M-1$. If $r_0 >0$, the entire population is eventually forced into the lowest fitness class, so that $f_{m}^{t \rightarrow \infty}=\delta_{m,M-1}$.    Conversely, if $r_0 <0$,  we maintain a stationary fitness distribution spread out over many fitness classes.   The phase diagram and steady-state solutions are illustrated in figure~\ref{fig:ratchetMF}.  A good order parameter for the phase transition (which will also be useful when we analyze range expansion simulations)  is the steady-state fraction of cells in the fittest class, $f_0^{t \rightarrow \infty}$.  If $f_{0}^{t \rightarrow \infty}(\mu,s)=0$, we are then  in the \textit{inactive} phase where the cells in the  $m=0$ class cannot survive at long times due to the deleterious mutations. Conversely, if $f_{0}^{t \rightarrow \infty}(\mu,s)>0$, we are in the \textit{active} phase and the  most fit $m=0$ cells constitute a non-zero fraction of the population, represented by the left edge of a Poisson distribution (see lower right inset of figure~\ref{fig:ratchetMF}).   This phase transition in the infinite population limit is not unique to our model.  For example, an analogous  fitness distribution collapse occurs in an infinite-population model of the evolution of DNA binding sequences \cite{gerlandhwa}. 

When $M \rightarrow \infty$, the order parameter is non-zero for any $\mu,s>0$, since $f_0^{t \rightarrow \infty} = e^{-\mu/s} >0 $,  and the phase transition disappears!     However, this behavior is an artifact of the limit of infinite population size.  If we had some finite population size $N$ and introduced the noise term back into equation~(\ref{eq:UCDPLang}), then a t\a transition can arise if  the number of individuals in the $m=0$ class, $n_0 \approx N e^{-\mu/s}$, is driven to zero via number fluctuations, leading to an irreversible loss of individuals in the highest $m=0$ fitness class, and a shift in the fitness distribution.  A succession of such shifts is known as Muller's ratchet in a well-mixed population \cite{ratchet2}.   The differences between Muller's ratchet and the error threshold in a well-mixed population has been discussed further in population genetics literature \cite{wagner}.  The lack of a transition for $M \rightarrow \infty$  arises because Muller's ratchet does not operate in a noise-free infinite, well-mixed population.  The finite-$M$ error threshold transition is the relevant mean-field description of the fitness collapse in the spatial  range expansions of interest here: it accurately describes range expansions above the upper-critical dimension $d=3+1$.  However, in biologically relevant dimensions, such as the $d=2+1$ and $d=1+1$ cases discussed in the introduction, the diffusion and noise terms in  equation~(\ref{eq:UCDPLang}) significantly modify the mean field result.

Consider the mean-field transition in more detail: When we approach the critical point from the active phase, $r_0 \rightarrow 0^-$, the expression $r_m=\mu-s(M-1-m)$ simplifies: $r_m \rightarrow sm>0$ for all $m>0$.  Thus, $r_m$ is positive, and all of the other classes $m>0$ are already in their inactive state.  Consequently, the cells with classes $m>0$ have no interesting critical dynamics of their own and inherit the critical scaling of the $m=0$ class via  the mutation term $\mu f_0$.  So, as $r_0 \rightarrow 0^-$, the steady-state fractions of  all the species  vanish according to $f_m^{t \rightarrow \infty} \sim |r_0|^{\beta_{\mathrm{MF}}^{(m)}}$ ($0 \leq m < M-1$), where the critical exponents are $\beta_{\mathrm{MF}}^{(m)}=1$.  In other words, the critical exponents assume the $m$-independent value  given by the mean-field DP  exponent governing the critical behavior of the $m=0$ class.  Thus, the densities of all fitness classes vanish in the same way within mean-field theory.

In $d=1+1$ and $d=2+1$-dimensional range expansions (and even for $d=d'+1$ with $d'\geq 3$), the noise and diffusion terms in equation~(\ref{eq:UCDPLang}) will renormalize the coefficients $r_m$ and $T_{m \ell}$.   These renormalizations lead us to consider general couplings  $r_m$ and $T_{m \ell}$,  which leads to a rich variety of possible critical behaviors.  For example, consider the $M \rightarrow \infty$ case.  Otwinowski and Krug argued that the transition to a fitness collapse  in a two-dimensional range expansion $(d=1+1)$ with $M \rightarrow \infty$ will fall into the same universality class as the interface growth models which exhibit multicritical, unidirectionally coupled directed percolation (UCDP) behavior \cite{spatialratchet, RougheningCoupledDP}.   The UCDP scaling regime is achieved in equation~(\ref{eq:UCDPLang}) when we fix $T_{m \ell} = T \delta_{m \ell}$ and  let $r_m = r \rightarrow 0^-$ for all $m \geq 0$ \cite{TauberCoupledDP,TauberCoupledDP2,RougheningCoupledDP}.  In this case, the exponents $\beta^{(m)}$ for each class $m$ are quite different!  In particular, by analyzing equation~(\ref{eq:fitrecursion}) with these new coefficients, one finds that  $f_m^{t \rightarrow \infty} \sim |r|^{\beta_{\mathrm{MF}}^{(m)}}$ as $r \rightarrow 0^-$, where $\beta_{\mathrm{MF}}^{(m)} = 1/2^m$ \cite{RougheningCoupledDP}.  Note that in this case, only the $m=0$ class fraction $f_0^{t \rightarrow \infty}$ has mean-field DP scaling.  The other fractions have  exponents $\beta_{\mathrm{MF}}^{(m)}$ that rapidly decay with $m$.  This is the mean-field version of the  unidirectionally coupled directed percolation universality class.    If we include the off-diagonal terms in $T_{m \ell}$, we may find universality classes different from UCDP. These off-diagonal terms represent competition between different classes and become more relevant in higher dimensions  \cite{AsymmCoupledDP}.    The lattice simulation results in the next section strongly suggest that the competition terms modify scaling in three-dimensional range expansions when $M \rightarrow \infty$.

    Another way to describe the fitness collapse transition is to study how the fractions $f_m(t)$ decay with time at the transition. Ideas from directed percolation theory lead us to expect that these fractions  vanish according to the power laws 
$f_m(t) \sim t^{-\delta^{(m)}}$ where  $\delta^{(m)}=\beta^{(m)}/\nu^{(m)}_{\parallel}$ \cite{NEQPTBook}.  Here, the exponents $\nu_{\parallel}^{(m)}$ are time-like correlation length exponents for each class $m$ (see next section for more details).  In the mean-field approximation, $\nu_{\parallel,\mathrm{MF}}^{(m)} = 1$ and   $\delta_{\mathrm{MF}}^{(m)} = \beta_{\mathrm{MF}}^{(m)}/\nu^{(m)}_{\parallel,\mathrm{MF}}  = 1$ for all $m$ in equation~(\ref{eq:UCDPLang}).  At the mean-field UCDP point, however,  $\delta_{\mathrm{MF}}^{(m)} = 1/2^m$, which decreases rapidly with $m$.  A priori, it is not clear which (if any) of these mean-field behaviors  describes the evolutionary dynamics of range expansions, since the diffusion and noise terms are expected to modify the critical behavior. However, the mean field results help map out the space of possibilities.  We will use simulations in the next section to calculate the exponents $\delta^{(m)}$ for two- and three- dimensional range expansions for the first few fitness classes $m$.

  In addition to changing the scaling exponents, the noise and diffusion terms in equation~(\ref{eq:UCDPLang}) shift the critical line in the $(s,\mu)$-plane relative to its mean-field counterpart $r_0 = \mu -s(M-1)=0 $.  We expect that the local genetic drift at the population frontier will  induce a transition even for $M \rightarrow \infty$.    In the Krug and Otwinowski model with $M \rightarrow \infty$, for example, the transition in two-dimensional range expansions occurs when $\mu/s^2 \approx 1$ \cite{spatialratchet}.   We shall see in the next section that there are important differences for three-dimensional range expansions.  Although there is a fitness collapse transition when $M \rightarrow \infty$, as in two-dimensional range expansions, and we find indications of multi-critical scaling similar to the UCDP universality class, the exponents we find are significantly different from expected UCDP scaling for $d=2+1$ \cite{TauberCoupledDP2}.  Also, the phase diagram shape is qualitatively different from the two-dimensional case.

\section{Lattice Simulation Results  \label{sec:dp}}

\begin{figure}[!ht]
\centering
\includegraphics[width=3.6in]{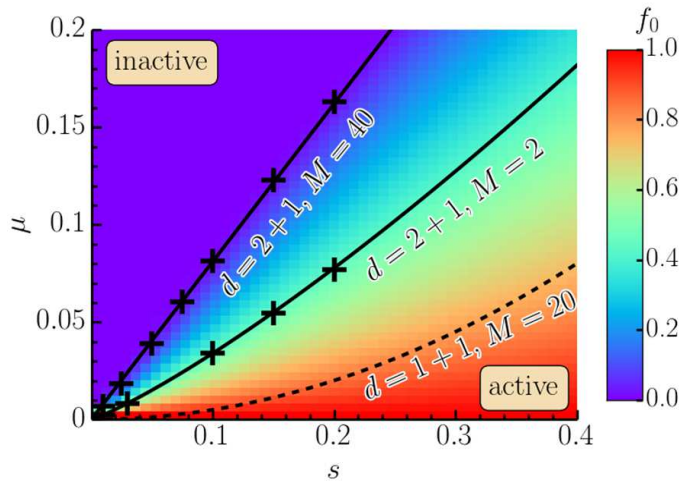}
\caption{ \label{fig:phaseDP} The phase diagram illustrating the transition from a stationary fitness distribution (active phase) to a collapsing fitness distribution (inactive phase) for three and two dimensional range expansions.  The heat map describes the case of a flat front with $M=40$ for $d=2+1$ dimensions, with  $M=40$ representing our approximation to the $M \rightarrow \infty$ limit. The color plot shows the  fraction of cells $f_0$  in the most favorable fitness class at $t=2000$ generations (averaged over 300 expansions)  for an initial  population frontier of $N=500^2$ cells (a triangular lattice 500 cells long and 500 cells wide)  all  in class 0. We believe that $t=2000$ closely approximates the steady-state value $f_{0}^{t \rightarrow \infty}$. We find precise locations (black crosses) at which the transitions for $M=40$ and $M=2$ occur by checking for dynamical critical scaling. The transition line for $M=40$ (solid line) is given approximately by  $\mu \approx 0.81 s$.  The $M=2$ transition has a  prominent logarithmic correction and can be fit by $\mu \approx 1.86 s/\ln (s/24) $.  The transition in $d=1+1$ dimensions (with $M=20$ effectively approximating the $M \rightarrow \infty$ limit)  has a very different shape, with $\mu \approx 0.5 s^2$, consistent with results of \cite{spatialratchet,MKNPRE}.   }
\end{figure}

 As discussed above, a natural order parameter for the fitness collapse transition is the infinite time limit of the fraction of cells in the fittest class, $m=0$, averaged over many range expansions with identical initial conditions.  This fraction, $f_{0}^{t \rightarrow \infty}(\mu,s)$, is then also averaged over the entire population front. The  initial condition is a population composed exclusively of cells in the most fit $m=0$ class.  We approximate $f_{0}^{t \rightarrow \infty}(\mu,s)$ in our simulations by  computing the fraction of $m=0$ cells  at some long time $t$ at which the system has already settled into a steady-state.      The corresponding phase diagrams in the $(s,\mu)$-parameter space for $d=1+1$ and $d=2+1$ are shown in figure~\ref{fig:phaseDP}.  Near the fitness collapse transition, the system takes longer and longer to settle into the steady-state due to critical slowing down.  However, at the resolution of the phase diagram of figure~\ref{fig:phaseDP}, the sampling time is long enough so that the corresponding corrections are negligible.  The phase diagram we find stops changing with increasing  $M$  for both $d=1+1$ and $d=2+1$ when $M \gtrsim 10$.    Hence, to understand the $M \rightarrow \infty$ case with less computational effort  we just look at  $M=40$ (as well as $M=2$) for $d=2+1$ in figure~\ref{fig:phaseDP}. For $d=1+1$, we take $M=20$ as a proxy for the limit $M \rightarrow \infty$.

  The phase transition line shape for two-dimensional expansions has been calculated previously \cite{MKNPRE,spatialratchet}.  It is given by $\mu \propto s^2$, where the constant of proportionality is model-dependent and will vary with $M$.  The particular scaling $\mu \propto s^2$, however, does not seem to change with $M$.  In our lattice model, we find that $\mu \approx 0.5 s^2$ for the large $M$ case, as shown by the dashed line in figure~\ref{fig:phaseDP}.   To calculate the shape of the transition line for three-dimensional (i.e. $d=2+1$) expansions, we start with the $M=2$ case and then check with simulations that only non-universal parameters  change with increasing $M$.

The transition line shape for three-dimensional expansions with $M=2$ may be calculated by analyzing the scaling near the special $s = \mu = 0$ point.   This point is described by the well-known voter model \cite{voter1} at which the two strains compete neutrally with no mutations.   A non-zero $s$ along the line $\mu=0$ introduces a \textit{bias} to the voter model and controls a phase transition.  Without mutations, our  initial condition of all $m=0$ cells is unchanged by the dynamics, and cannot be used to define an order parameter.  Instead, we take as our $\mu=0$ order parameter the survival probability of $m=0$  cells starting from an initial population frontier of  one $m=0$ cell (at the origin, say) surrounded by  $m=1$ cells. For $s < 0$, the $m=0$ cells engendered by this isolated mutant always die out (inactive phase), while for $s > 0$ there is a non-zero survival probability (active phase).    The critical point itself is at $s=0$.  Note that the voter model has its own universality class, different from directed percolation  \cite{voter2,NEQPTBook}.

 A non-zero mutation rate $\mu>0$ pushes the system out of the voter model class.  The corresponding cross-over scaling description has been calculated previously using field-theoretic techniques  \cite{CDPFT}.  The transition line shape, derived from cross-over scaling, has the form $\mu = A s/|\ln( s/B) |$. Both the constant of proportionality, $A$, and the constant $B$ inside the logarithm are non-universal. A similar transition line shape is found for the onset of mutualism in three-dimensional range expansions \cite{MLDRNPRL}.  We find these non-universal parameters by fitting to simulation data, as shown in figure~\ref{fig:phaseDP}.  This cross-over scaling shape works well for $M=2$, but the logarithmic correction is not as evident in the large $M$  case ($B$ is much larger than typical values of $s$).  Hence, we fit the line shape to $\mu = As$ in figure~\ref{fig:phaseDP}, with $A$ as our single fitting parameter.     This linear shape works extremely well (see $M=40$, $d=2+1$ transition line in figure~\ref{fig:phaseDP}).

\begin{figure}[!ht]
\centering
\includegraphics[width=4.5in]{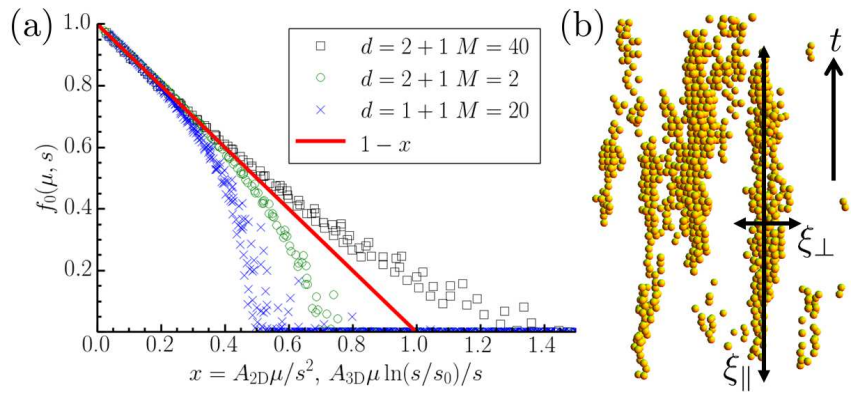}
\caption{ (a) The average steady-state fraction of the most fit 0-class cells, $f_{0}(\mu,s)$, for many different values of $\mu$ and $s$.  For $d=2+1$ and $M=40$, the data set is the same as in figure~\ref{fig:phaseDP}(a). For $d=1+1$, we evaluated $f_0(\mu,s)$ at $t=10^4$ generations (averaged over 2000 runs) for a frontier of size $N=2000$ cells and $M=20$ .  We also included the $d=2+1$,  $M=2$ case for comparison ($N=500^2$ cells, $t=2000$ generations, 128 runs).  Upon using the scaling variables in equation~(\ref{eq:ssfrac}), the data points collapse onto a single curve near  the region $f_0(\mu,s) \approx 1$ that describes the active phase.  For the $d=2+1$ cases, we find   $A_{\mathrm{3D}} \approx 0.3 $, $s_0 \approx 40 $ for $M=40$ and $A_{\mathrm{3D}} \approx 0.46$, $s_0 \approx 11$ for $M=2$.  For $d=1+1$, we find $A_{\mathrm{2D}} \approx 1.05$.   (b) A snapshot of all the $m=1$ clusters with $M=5$ in a section of a $d=2+1$ range expansion deep in the active phase ($s=0.08$, $\mu=0.0008$). These clusters an average width $\xi_{\perp}$ and length $\xi_{\parallel}$, illustrated schematically with the arrows.   \label{fig:ssfraccollapse}  }
\end{figure}

Scaling near the voter model point also determines the behavior of  $f_{0}^{t \rightarrow \infty}(\mu,s)$ deep in the active phase, when $f_{0}^{t \rightarrow \infty}(\mu,s)$ is close to 1.  In a well-mixed population, this means $\mu/s \ll M-1$ and we are far away from the  transition.  As discussed in  section~\ref{sec:meanfield},  $f_0^{t \rightarrow \infty} \approx e^{-\mu/s} \approx 1-\mu/s$ in this case. In $d=1+1$ dimensions, $f_0^{t \rightarrow \infty}$ may be calculated by mapping the boundaries of localized patches of less fit ($m>0$) strains to random walks  \cite{MKNPRE,nelsonhallatschek,spatialratchet}.  This mapping is essentially the same one employed to study clusters in the $d=1+1$ voter model.  However, the steady-state fraction $f_0^{t \rightarrow \infty}$ now depends on the scaling combination  $\mu/s^2$, instead of $\mu/s$.     The simulation results for $f_0^{t \rightarrow \infty}(\mu,s)$ are presented in figure~\ref{fig:ssfraccollapse}(a) for $d=1+1$ and $d=2+1$.  In $d=2+1$,  the mutant cluster shapes are quite complicated, as illustrated in figure~\ref{fig:ssfraccollapse}(b).  A simple mapping to random walkers is not possible in this case.  However, we will now make a scaling argument for $f_0^{t \rightarrow \infty}$ that exploits exact voter-model results for  $d=2+1$.

    Consider that deep in the active phase, there will be  finite-sized connected clusters of species with fitness classes $m>0$, and in class $m=1$ in particular.   The rest of the population will be in the fittest $m=0$ class. Provided clusters of cells with $m=1$ are so dilute that they do not collide, we can treat each cluster independently.  The $m=0$ state, in this case, acts as the \textit{absorbing} state for each cluster.  Note that this is backwards from the analysis of the fitness collapse transition, where we treat the $m=0$ state as the ``active'' state.  The non-interacting $m=1$ clusters  with some finite average width $\xi_{\perp}$ and length $\xi_{\parallel}$ are shown  in figure~\ref{fig:ssfraccollapse}(b).    If the $m=0$ class has some selective advantage coefficient $s>0$, the $m=1$ clusters will have a relative selective \emph{disadvantage} $-s < 0$.     So, the $m=1$ clusters  will be governed by voter model scaling in the \textit{inactive} state.   They will  contain some average number of cells $\langle n \rangle_s$, which will depend on $s$. Since the clusters do not collide, we can approximate $f_0^{t \rightarrow \infty}$ as
\begin{equation}
f_0^{t \rightarrow \infty}(\mu,s) \approx 1 - \frac{\mu \langle n \rangle_s}{N_{\mathrm{fr.}}}, \label{eq:steadystate3d}
\end{equation} 
where $N_{\mathrm{fr.}}$ is the total number of cells at the population frontier.  Also, since  the cell populations with  $m>1$ will be very small, the value of $f_0^{t \rightarrow \infty}(\mu,s)$ should not vary much with $M$.

 We now estimate  how the average number of cells $\langle n \rangle_s$ in an $m=1$ cluster scales with $s$.    Consider a cluster formed at the origin $\mathbf{r}=0$ at time $t=0$.  We may calculate the pair connectivity function $\Upsilon(|\mathbf{r}|,t; |s|)$, which is the probability of finding a cell in the cluster a distance $|\mathbf{r}|$ from the origin along the population frontier at time $t$ \cite{NEQPTBook}.  The scaling behavior of $\Upsilon(|\mathbf{r}|,t;|s|)$ near the phase transition at $s=0$ may be extracted from the general scaling relations derived in \cite{CDPFT}.  At $d=2+1$, we are at the upper critical dimension of the voter model and there are logarithmic corrections.  Using the general scaling relations in \cite{CDPFT}, we find that  $\langle n \rangle_s$ satisfies
\begin{eqnarray}
\fl \langle n \rangle_s & \simeq \frac{1}{a^2 \tau_g} \int \mathrm{d}^2 \mathbf{r}\,\mathrm{d} t \, \Upsilon(|\mathbf{r}|,t; |s|) \\ \fl & \simeq   \frac{1}{a^2 \tau_g \lambda |\ln (\lambda/ \lambda_0)|} \int \mathrm{d}^2 \mathbf{r}\,\mathrm{d} t \, \Upsilon \left[ \frac{|\mathbf{r}|}{ \sqrt{\lambda | \ln (\lambda/ \lambda_0 )|}} , \frac{t}{\lambda |\ln( \lambda/ \lambda_0)|}; \lambda |s| \right],  \label{eq:avgnscaling}
\end{eqnarray}
where $a$ is a cell diameter, $\tau_g$ a generation time,  $\lambda$ an arbitrary scale factor, and $\lambda_0$ is some non-universal constant. The two-dimensional integral over $\mathbf{r}$ ranges over the entire population frontier.  This integral is of order the frontier area, which is approximately $N_{\mathrm{fr.}} a^2$.  The  $|s|$-dependence of $\langle n \rangle_s$ may now be extracted from equation~(\ref{eq:avgnscaling}) using a judicious choice of $\lambda=1/s$ and an appropriate rescaling of both $\mathbf{r}$ and $t$.  After these manipulations, we find that in the inactive phase:
\begin{equation}
\langle n \rangle_s \simeq A_{\mathrm{3D}} N_{\mathrm{fr.}} \left| \frac{  \ln (s/s_0)}{s} \right|,
\end{equation}
where $A_{\mathrm{3D}}$ is a constant proportional to the residual integration in equation~(\ref{eq:avgnscaling}) and $s_0=\lambda_0^{-1}$. Upon returning to our expression for $f_0$ in equation~(\ref{eq:steadystate3d}), we find that
\begin{equation}
f_0^{t \rightarrow \infty} \approx 1 - \frac{ A_{\mathrm{3D}} \mu  }{s} \, |\ln (s/s_0)|.
\end{equation}

Upon combining this analysis with results for well-mixed and $d=1+1$ dimensional systems, we expect the following small $\mu$ behaviors in the active phase:
\begin{equation}
f_{0}^{t \rightarrow \infty}(\mu,s) \approx \cases {
 1- \frac{\mu}{s} & well-mixed    \\
 1 - A_{\mathrm{2D}}\, \frac{\mu}{s^2} & $d = 1+1$ \\
 1 - A_{\mathrm{3D}}\, \frac{\mu}{s} \,  |\ln (s/s_0)| & $d = 2+1$
}, \label{eq:ssfrac}
\end{equation}
where $A_{\mathrm{2D},\mathrm{3D}}$  and $s_0$ are non-universal and depend on the details of the models.   We check the validity of equation~(\ref{eq:ssfrac})  for $d=1+1$ and $d=2+1$ for small $\mu$ with simulations in figure~\ref{fig:ssfraccollapse}. For our model, we find $A_{\mathrm{2D}} \approx 0.5$, $A_{\mathrm{3D}} \approx 0.3$, and $s_0 \approx 40$.   These parameters yield good data collapses in the active phase, i.e., for values of $x \lesssim 0.2$ in figure~\ref{fig:ssfraccollapse}. As we move away from the strongly active phase region,  the approximations in equation~(\ref{eq:ssfrac}) start to fail due to cell cluster collisions. Eventually, we reach the fitness collapse regime which we now investigate in more detail.

\begin{figure}[!ht]
\centering
\includegraphics[width=4.5in]{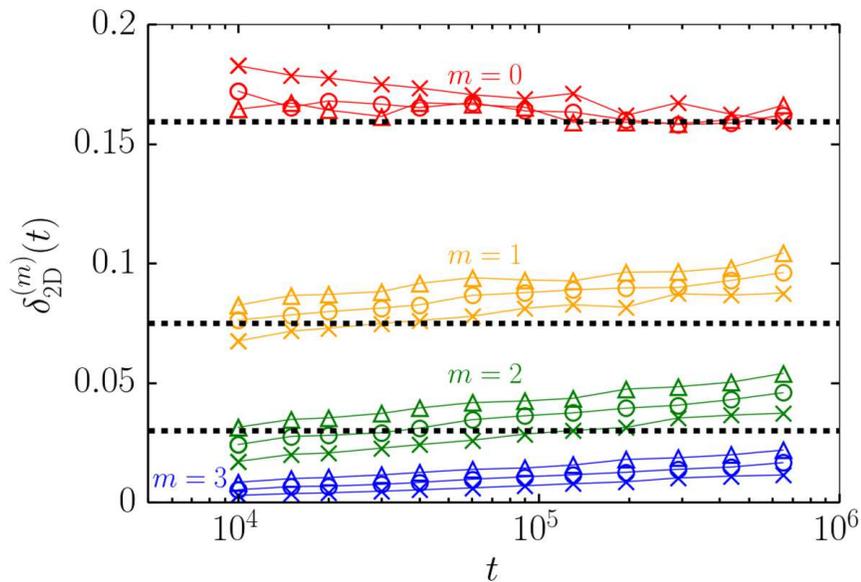}
\caption{ \label{fig:effexp2d}  Effective exponents at the fitness collapse transition versus time $t$ in a two-dimensional range expansion evolved for $t$ generations with a one-dimensional frontier of  $N=10^4$  cells and an initial population entirely in the $m=0$ class.  We tune $\mu$ to the fitness collapse transition for three different values of $s$: $s=0.05$ (crosses) $s=0.1$ (circles), and $s=0.2$ (triangles).  The lines connecting these points are to guide the eye. The black dashed lines show the expected UCDP exponents  \cite{TauberCoupledDP2}.  Previously calculated exponent values for $m=3$ were not available.}
\end{figure}

A signature of the phase transition, when $\mu$ is tuned upward to reach fitness collapse, is the power-law decay of the fitness class fractions with time:
\begin{equation}
f_m \sim t^{-\delta^{(m)}},
\end{equation}
where $\delta^{(m)}=\beta^{(m)}/\nu_{\parallel}^{(m)}$ represents the family of critical exponents introduced in the previous section.  A good technique for finding these exponents from simulations is to calculate the time-dependent effective exponent $\delta^{(m)}(t)$:
\begin{equation}
\delta^{(m)}(t=t_i)= \frac{\ln [f_m(t_i)/f_m(t_{i-1})]}{\ln[ t_i/t_{i-1}]}, \label{eq:effexpdelta}
\end{equation}
where the  $\{t_i\}$ are the sampled times in our simulation, chosen so that $t_{i}/t_{i-1} \approx 2$.  The  exponent may be calculated by estimating the steady-state value of $\delta^{(m)}(t)$ at long times.  The error may be estimated by looking at the fluctuations of $\delta^{(m)}(t)$.  This technique works well for the $m=0$ class, but we must introduce a  modification for $m>0$.  Just as in the interface growth models \cite{RougheningCoupledDP}, the exponents converge faster if we look at the integrated fractions $F_m \equiv \sum_{\ell = 0}^m f_{\ell}$.  Then, since $f_{m-1} \ll f_m$ near the fitness collapse transition, we expect that $F_m \sim  t^{-\delta^{(m)}}$.  We use these integrated fractions $F_m$, instead of $f_m$, to calculate $\delta^{(m)}$ via the effective exponent technique.  We do this for both $d=1+1$ and $d=2+1$.

 The effective exponents $\delta^{(m)}_{\mathrm{2D}}(t)$ for $d=1+1$ are shown in figure~\ref{fig:effexp2d}, calculated using the multiple color generalization of the Domany-Kinzel model discussed above \cite{domany,MKNPRE}.  We see that the exponents $\delta_{\mathrm{2D}}^{(m)}$ decrease with increasing $m$, as we would expect at the multicritical UCDP point discussed in the previous section.  The effective exponent results are compared to exponents reported for UCDP in the literature (dashed lines in figure~\ref{fig:effexp2d}) \cite{TauberCoupledDP2} .  Our results [$\delta^{(0)}_{\mathrm{2D}} \approx 0.164(6)$, $\delta^{(1)}_{\mathrm{2D}} \approx 0.087(12)$, $\delta^{(2)}_{\mathrm{2D}} \approx 0.035(13)$] are consistent with the UCDP exponents for $m=0,1,2$.  The $m=3$ exponent does not appear to have a reported value and we compute $\delta^{(3)}_{\mathrm{2D}}(t) \approx 0.012(8)$.  Hence, the fitness collapse transition for two-dimensional range expansions appears to be governed by UCDP, the universality class of the interface models discussed in \cite{RougheningCoupledDP, TauberCoupledDP, TauberCoupledDP2}. This is an important conclusion, because it means that the off-diagonal competition terms $T_{m \ell}$ ($m \neq \ell$) in equation~(\ref{eq:UCDPLang}) are evidently irrelevant in $d=1+1$ dimensions. We shall see that the situation in three-dimensional expansions is quite different.

In the inactive phase, there will be a local speed $V$ associated with the collapsing fitness distribution.   This speed is the analogue of the Muller's ratchet ``clicking'' rate as the fitness diminishes in well-mixed populations. As discussed in \cite{spatialratchet},  the exponents $\nu_{\parallel}^{(m)}$ govern the speed $V$.  Previous results for $d=1+1$ suggest that the exponents $\nu_{\parallel}^{(m)}$ in the UCDP class do not  vary with $m$ and are all equal to the DP value $\nu_{\parallel}^{(m)} = \nu_{\parallel} \approx   1.733847(6)$ \cite{NEQPTBook, RougheningCoupledDP}. So, in two-dimensional expansions, we expect that the fitness distribution speed is  $V \sim |r|^{\nu_{\parallel}}$, where $r= \mu/s^2-\mu_c/s_c^2$ is the distance away from the critical point [with $(s_c,\mu_c)$ a critical point on the line of DP transitions].      Finally, note that there is a slow, upward drift in the $m>0$ effective exponents $\delta^{(m)}_{\mathrm{2D}}(t)$ as $t$ increases in figure~\ref{fig:effexp2d}.  This effect was noticed in other UCDP models \cite{TauberCoupledDP2}. Hence, it is possible that the multicritical regime is only relevant at intermediate times and that  all of the exponents eventually approach the DP value reached by the $m=0$ effective exponent, as one might expect from mean field theory.  Nevertheless, the multicritical behavior is clearly important for a wide range of times as the population evolves.

\begin{figure}[!ht]
\centering
\includegraphics[width=3.3in]{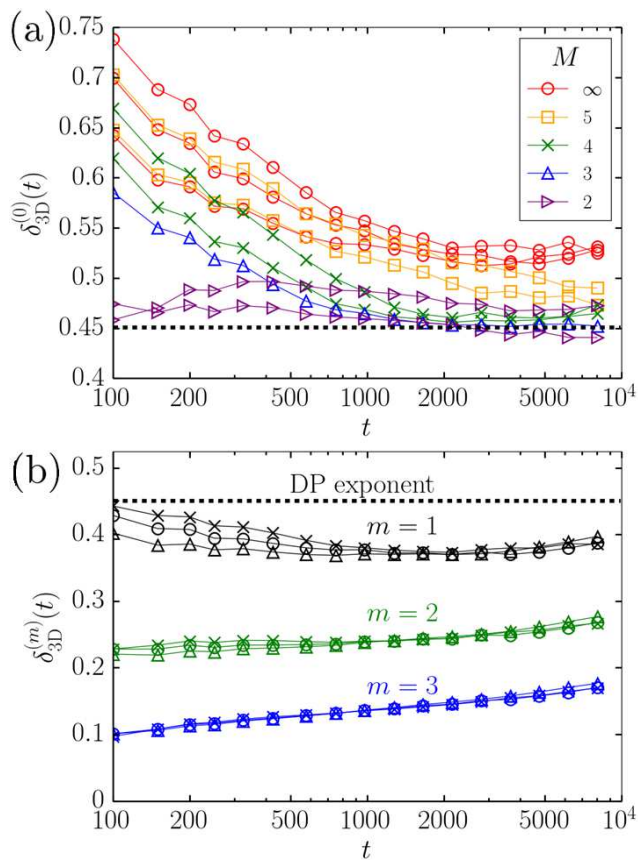}
\caption{ \label{fig:effexp3d} (a) The effective exponent $\delta^{(0)}_{\mathrm{3D}}(t)$ (defined as in equation~(\ref{eq:effexpdelta})) describing the temporal decay of the fraction $f_0$  of 0-class cells  in a three-dimensional range expansion with a frontier of $N=700^2$ cells at the fitness collapse transition  (averaged over at least 1000 runs).    The range expansions have different values of $M$ and an initial population of cells all in the  most fit $m=0$ class.  The different sets of points (with connecting lines to guide the eye) at the same $M$ correspond to different positions along the line of phase transitions shown in figure~\ref{fig:phaseDP}. For the $M \rightarrow \infty$, the number of mutations a cell may acquire during a simulation run is unbounded.  (b) Effective exponents for larger values of $m$ for the $M \rightarrow \infty$ case.  As in the $d=1+1$ case (figure~\ref{fig:effexp2d}), the exponents decrease with increasing $m$, indicating a multicritical scaling regime.  We   tune $\mu$ to the fitness collapse transition for three values of $s$: $s=0.075$ (crosses) $s=0.1$ (circles), and $s=0.15$ (triangles).  In both (a) and (b), the dashed line indicates the expected directed percolation (DP) exponent value $\delta_{\mathrm{3D}}^{(0)} \approx 0.451$ \cite{NEQPTBook}.  }
\end{figure}

We now analyze the fitness collapse transition in three-dimensional range expansions.   We've seen that the scaling combination of $s$ and $\mu$ that determines the shape of the phase transition line is $x=\mu/s$ (up to logarithmic corrections).  Hence, there will be some critical value $x_c$ describing the transition line in the $(s,\mu)$ plane. The parameter  $r \equiv x-x_c$ will measure the distance away from the transition.    For $M=2$, we expect the 0-class fraction at $r=0$  to decay with the DP exponent $\delta^{(0)}_{\mathrm{3D}} \approx 0.4505(10)$ \cite{NEQPTBook}.    The results for $\delta^{(0)}_{\mathrm{3D}}(t)$ for various $M$ are shown in figure~\ref{fig:effexp3d}(a).   As expected, the effective exponents $\delta^{(0)}_{\mathrm{3D}}(t)$ for small $M=2,3,4$  approach the DP value (dashed line).   However, as $M \rightarrow \infty$, we find a significant deviation from DP scaling.  The deviation seems to occur at around $M \gtrsim 5$.   We find a scaling exponent $\delta_{\mathrm{3D}}^{(0)} \approx 0.53(2)$ for $M \rightarrow \infty$ that is higher than the expected DP value [circles in figure~\ref{fig:effexp3d}(a)].  This value does not appear to change along the transition line.  Thus, our exponent deviates from what is observed in the interface growth models.  Note that in the Langevin equation proposed for the interface model \cite{RougheningCoupledDP}, the $0$-class dynamics obey the DP equation \cite{NEQPTBook}.  Our equation, the $m=0$ case in equation~(\ref{eq:UCDPLang}),  is  different because it includes off-diagonal terms in the matrix $T_{m \ell}$.  These competition terms might be responsible for the deviation from DP scaling  at large $M$.   This deviation is also expected from mean-field theory, where adding the competition terms can increase the exponent $\delta^{(0)}$ by a factor of 2 \cite{AsymmCoupledDP}.

 We also check to see if there is multicritical behavior in this model when $M \rightarrow \infty$. We do find that when $M \rightarrow \infty$, the higher order exponents $\delta^{(m)}_{\mathrm{3D}}$ decrease with increasing $m$, as shown in figure~\ref{fig:effexp3d}(b).   The exponent values we find, however, are quite different from the expected UCDP values.  We find $\delta^{(0)}_{\mathrm{3D}} \approx 0.53(2)$, $\delta^{(1)}_{\mathrm{3D}} \approx 0.38(2)$, $\delta^{(2)}_{\mathrm{3D}} \approx 0.24(2)$, and $\delta^{(3)}_{\mathrm{3D}} \approx 0.14(3)$, compared to previously calculated values $\delta^{(0)}_{\mathrm{3D}} \approx 0.46(2) $, $\delta^{(1)}_{\mathrm{3D}} \approx 0.26(3) $, and $\delta^{(2)}_{\mathrm{3D}} \approx 0.13(3)$ for UCDP with $d=2+1$ \cite{TauberCoupledDP2}.   We also see an upward shift in the exponents over time, just as in the two-dimensional range expansion case in figure~\ref{fig:effexp2d}. Hence, the multicritical regime might be transient.  Nevertheless, the $m=0$ effective exponent, $\delta^{(0)}_{\mathrm{3D}}(t)$, appears to settle to a fixed value. So, there is clearly some interesting deviation from regular DP behavior as we increase the number of species $M$.    Assuming the exponents $\nu_{\parallel}^{(m)}$  for various $m$ do not deviate  from the directed percolation value, the speed of the fitness wave associated with the fitness collapse should be governed by $d=2+1$-dimensional DP \cite{NEQPTBook}:
\begin{equation}
V \sim \left| r \right|^{\nu_{\parallel}} \mbox{ \quad with \quad} \nu_{\parallel} \approx 1.2950(60).
\end{equation}
Our simulation results are consistent with this scaling of the speed $V$ (data not shown), but a thorough check of the $v_{\parallel}^{(m)}$ exponents is beyond the scope of this paper. More extensive simulations would be necessary to verify that our model falls into a class distinct from DP and is not exhibiting a  long-lived transient.  For biological applications, however, our analysis is relevant because the multicritical behavior can have measurable effects over thousands of generations.

\section{Effects of Inflation \label{sec:inflation}}

 \begin{figure}[!ht]
\centering
\includegraphics[width=3.8in]{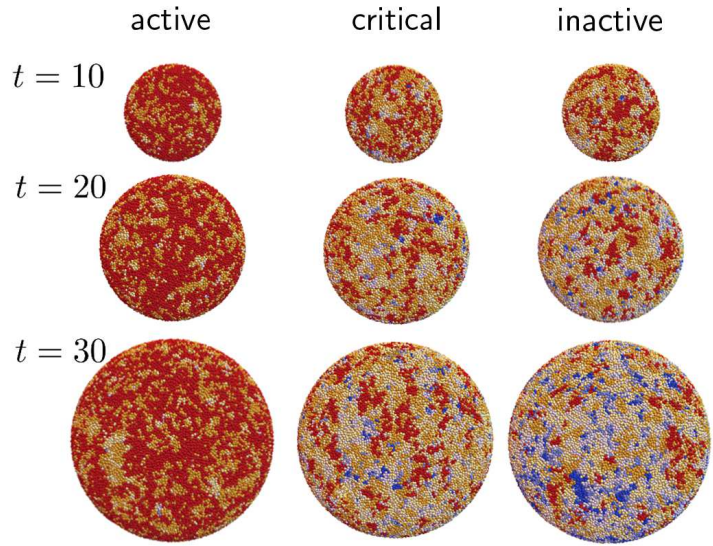}
\caption{ \label{fig:phaseDPSphere}  Spherical range expansions also exhibit a transition, but certain features are smoothed out, similarly to a finite size effect.   At different times $t$ (measured in generations), we show the surfaces  of growing spherical populations with $M=6$ and initial radii  of about ten cell diameters, $R_0\approx 10 a$.  The initial populations have all $0$-class (red) cells. The cells turn more blue as they acquire more mutations.  We show the population in the active region ($\mu=0.05$, $s=0.2$), inactive region ($\mu=s=0.1$), and near criticality ($\mu=0.1$, $s=0.14$).  }
\end{figure} 
 
 Spherical range expansions also exhibit a fitness collapse, as illustrated in  figure~\ref{fig:phaseDPSphere}.  Deep in the active phase, the inflationary nature of the population frontier will be largely irrelevant because, as discussed in the previous section, the evolutionary dynamics will be governed by small clusters with $m>0$, which are insensitive to the front inflation.  Similarly, large enough mutation rates will eventually extinguish the $0$-class individuals in the inactive phase.  Inflating frontiers will most strongly influence the evolution near the onset of fitness collapse.  As argued in an earlier study of two-dimensional, circular expansions \cite{MKNPRE}, inflation will causally disconnect portions of the population and prevent correlations from propagating along the frontier.  There is still an abrupt population collapse, but certain critical properties are modified, similar to a finite size effect.

    For simplicity, we study the modifications due to inflation for spherical expansions just for the $M=2$ case.   The comparison between inflating and non-inflating frontiers for $M=2$ is particularly informative, because we know that the non-inflating expansions are characterized by a genuine DP transition.  Higher values of $M$ will also exhibit a transition, as illustrated in figure~\ref{fig:phaseDPSphere}.  We expect the effects of inflation to be similar for increasing $M$ (and the $M \rightarrow \infty$ case in particular), but a direct comparison would require a better characterization of the universality class of  non-inflating, three-dimensional range expansions.  As discussed in the previous section, the $M \rightarrow \infty$ case appears to violate DP scaling, but a thorough investigation of this potentially new universality class is beyond the scope of this paper.

An important confounding factor is the presence of a finite system size.  Note that in the spherical expansion illustrated in figure~\ref{fig:simintro}(b) and figure~\ref{fig:phaseDPSphere}, the genetic patterns evolve on a finite spherical surface.  Simplifications arise for flat range expansions like those in figure~\ref{fig:simintro}(a), where an arbitrarily large frontier area mitigates finite size effects.  To disentangle the effects of the finite surface area and the effects of inflation, we will compare inflating spherical expansions, as in figure~\ref{fig:phaseDPSphere}, to populations at the surface of \textit{treadmilling} spheres of some fixed radius $R_0$.  The cells at the frontier of these fixed radius spheres will divide and displace each other over time, \ thus creating a dynamical ``treadmilling'' effect.  As discussed in the introduction,   such a treadmilling sphere can model  an avascular tumor which turns over cells at its surface but does not expand due to apoptosis at its center and pressure from the surrounding tissue \cite{treadmill1}. The treadmilling sphere simulations are performed by growing an initial spherical population of radius $R_0$ out to radius $R_0+n a$, with $a$ a cell diameter and $n \approx 2$.  This ``forward sweep'' has the same update rules described in section~\ref{sec:model} for spherical range expansions.  After the sweep, the outermost shell of the population is treated as an initial condition to evolve the population \textit{backwards} to radius $R_0-na$, using the time-reversed version of the update rule discussed in section~\ref{sec:model}.  Specifically, cells compete to divide into empty sites chosen in the reversed order from the forward sweep. The forward and backward sweeps are repeated many times, creating a dividing population of a fixed size at a distance $R_0$ from the sphere center.  For more details, see \cite{MLSphere}. Note that our results for how the finite size of the population influences the evolutionary dynamics at the fitness collapse transition should be insensitive to the particular details of how the treadmilling population is established.

For $M=2$, an important ``rapidity reversal'' symmetry exists at the DP transition \cite{Hinrichsen}. This symmetry affects the survival probability $P(t)$  of a mutant cluster formed from an initial condition with a  single $m=0$ class cell surrounded by cells in the $m=1$ class. Rapidity reversal symmetry insures that, at long times, $P(t)$ is proportional to the mutation-driven decay of the fraction $f_0(t)$ of $0$-class cells starting from an initial population entirely in class   $m=0$. In particular, $P(t) \approx \kappa f_0(t) \sim t^{-\delta^{(0)}}$ at the transition, where $\delta^{(0)} \approx 0.451$  is the DP critical exponent \cite{Hinrichsen}.  The constant of proportionality $\kappa$ depends on $\mu$ and approaches zero as $\kappa \sim \sqrt{\mu}$  when $\mu \rightarrow 0$. In inflating, two-dimensional (circular) range expansions, this rapidity reversal symmetry is \textit{broken} after the cross-over time $t^* = R_0/v$  (where $v$ is the front speed) because the wandering of the mutant cluster boundaries gets overwhelmed by the inflating perimeter \cite{MKNPRE}.   We expect a similar symmetry violation with respect to rapidity reversal in three-dimensional range expansions.

\begin{figure}[!ht]
\centering
\includegraphics[width=3.7in]{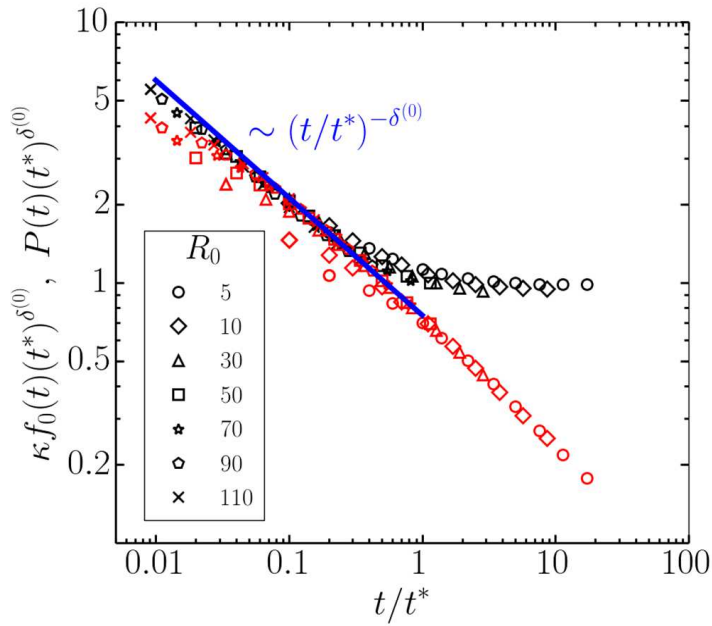}
\caption{ \label{fig:rapiditybreak} Rapidity reversal symmetry in spherical range expansions spoiled by inflation near the fitness collapse transition  ($\mu = 0.1$, $s = 0.2586$). The red symbols correspond to the decaying fraction of $0$-class cells $f_0(t)$ for an initial population entirely in the $m=0$ class. The black symbols show the survival probability $P(t)$ of a cluster generated from a single $0$-class cell. Both $f_0(t)$ and $P(t)$ are averaged over at least $3 \times 10^4$ runs.  We find $\kappa \approx 0.6$ by adjusting the data collapses until these two quantities overlap at early times. Note that these quantities decay in a similar way for  $t \ll t^*$, consistent with DP scaling, indicated by the solid blue line. At late times, however, the survival probability approaches a non-zero constant $P(t \rightarrow \infty)\sim (t^*/\tau_g)^{-\delta^{(0)}}$, where $\delta^{(0)}  \approx 0.451$ \cite{Hinrichsen}, while $f_0(t)$ continues to decay.  }
\end{figure}

The rapidity reversal results are shown in figure~\ref{fig:rapiditybreak}.  We indeed find that $P(t) \approx \kappa f_0(t)$, with $\kappa \approx 0.6$, at early times such that $ t \ll t^*$.  However, if a mutant cluster survives past the cross-over time $t^*$, the cluster  will inflate along with the frontier, thus allowing it to survive indefinitely with some limiting survival probability $P_{\infty}=P(t \rightarrow \infty)$.  Our simulations indicate that $P(t)$  saturates after time $t > t^*$,  approaching a constant given approximately   by $ (t^*/\tau_g)^{-\delta^{(0)}}$  (black points in figure~\ref{fig:rapiditybreak}).  Since $t^* = R_0/v$, we estimate that  $P_{\infty} \sim (t^*/\tau_g)^{-\delta^{(0)}}= (a/R_0)^{\delta^{(0)}}$ near the fitness collapse transition.  Note that this result is dramatically different from a non-inflating, three-dimensional range expansion, for which we find $P_{\infty}=0$ at the transition.  This survival probability enhancement is also present at the voter model point, where  $P_{\infty} \sim a/R_0$ for spherical expansions \cite{MLSphere}.  Conversely, the fraction $f_0(t)$ of $0$-class cells will continue to decrease at the transition when we have an initial population of $m=0$ cells.  So, just as in two-dimensional circular range expansions, rapidity reversal is broken after time $t^*$ in three-dimensional spherical range expansions.

\begin{figure}[!ht]
\centering
\includegraphics[width=3.4in]{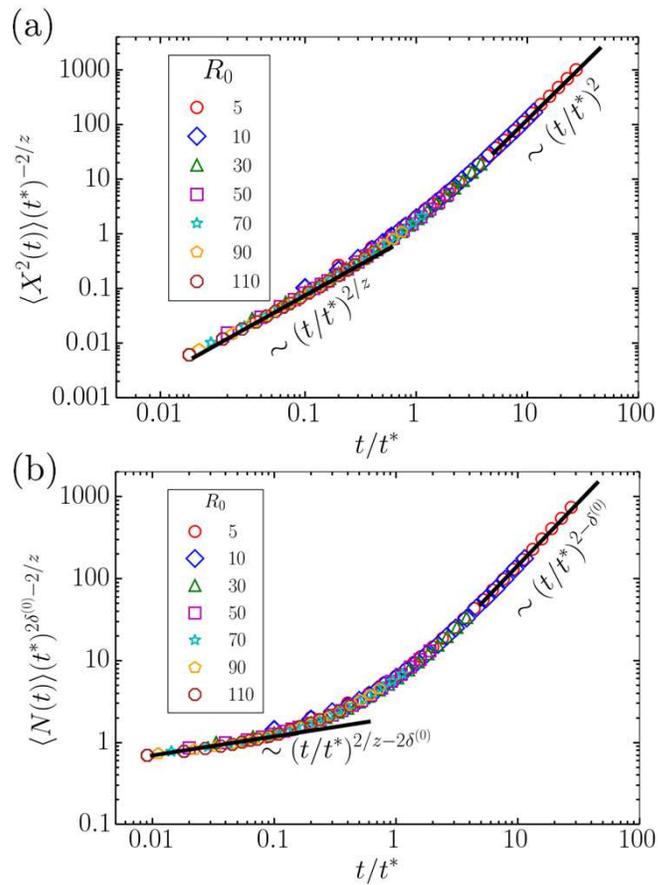}
\caption{\label{fig:singleseedDPscaling} Scaling of the cluster formed from a single mutant $m=0$ cell near the  transition.  The solid lines in both panels show the expected DP scaling for $t \ll t^*$ and inflationary  scaling for $t \gg t^*$, where $z \approx 1.7660(16) $ is the dynamic exponent for DP \cite{NEQPTBook}. All results are averaged over at least $3 \times 10^4$ runs.  In (a), we show the average squared spread $\langle X^2(t)\rangle$ of all clusters that survive at least until time $t$.  When $t \gg t^*$, inflation takes over and the mean cluster spread locks into the linear increase of the inflating sphere radius in time.  In (b), we track the number $N(t)$ of $m=0$ cells in the cluster at time $t$.   At  long times $t \gg t^*$, the cluster spread increases linearly in time, but the cell fraction inside the cluster decreases as $t^{- \delta^{(0)}}$, yielding the long time scaling behavior indicated by the solid line.   }
\end{figure}

We now study other properties of the cluster formed from a single $m=0$ mutant cell to better understand the effects of inflation. Two key quantities are the average squared cluster spread $\langle X^2(t) \rangle$ and the average number of $0$-class cells  $\langle N_0(t) \rangle$ in all surviving $0$-class  clusters at time $t$.  The spread $X(t)$ is the arc length between the position of the initial $m=0$ cell and a $m=0$ cell in the surviving cluster at time $t$. We average $X^2(t)$ over all $m=0$ cells at the frontier at time $t$ and over many simulation runs.      In the inflationary regime $t>t^*$, the area covered by the cluster increases approximately quadratically in  time $t$, due to inflation.  Inflation thus  leads to the long time scaling $\langle X^2(t) \rangle \sim t^2$ shown in figure~\ref{fig:singleseedDPscaling}(a).  Inside this quadratically increasing area, a critical directed percolation process occurs, with a decaying mutant fraction $f_0(t) \sim t^{-\delta^{(0)}}$. Upon combining the quadratic area scaling with the mutant fraction decay scaling, we find that $\langle N_0(t) \rangle \sim (t/t^*)^{2-\delta^{(0)}}$ for  $t \gg t^*$, as shown by the solid black line in figure~\ref{fig:singleseedDPscaling}(b).   These results indicate that the large scale features of the cluster, such as its spatial spread at time $t$, are dictated by the inflating population frontier.  Local features such as the local $0$-class fraction decay, however, retain their DP properties.

\begin{figure}[!ht]
\centering
\includegraphics[width=3.3in]{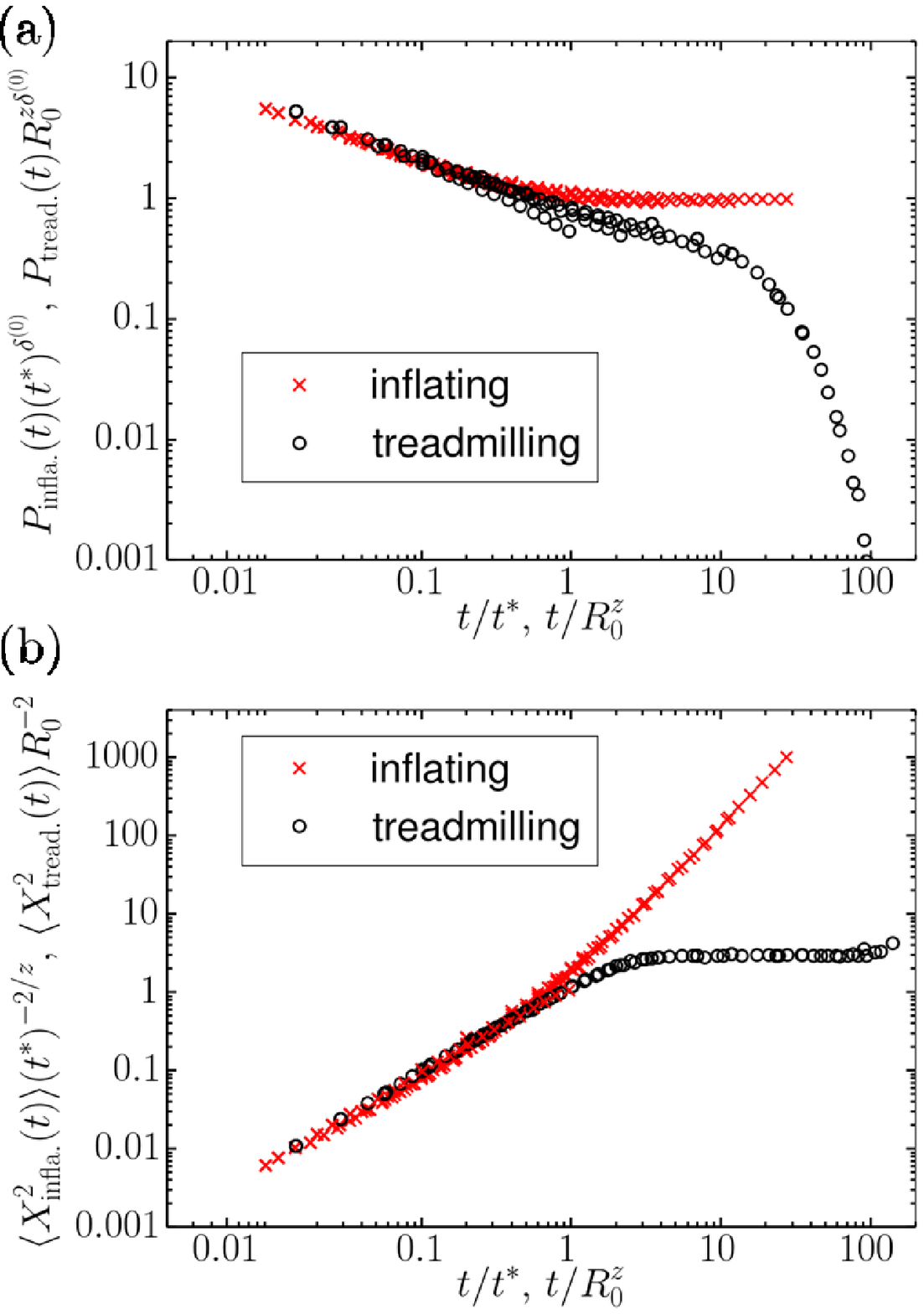}
\caption{\label{fig:treadmillsingleseed} A comparison of (a)  the survival probability and  (b) the average cluster spread for treadmilling  and inflating  spherical populations. The red crosses for the inflating expansions in (a) and (b) are derived from the same data as in figure~\ref{fig:rapiditybreak} and figure~\ref{fig:singleseedDPscaling}(b), respectively, with again the initial radii $R_0=5,10,30,50,70,90,110$. The black circles are a data collapse of many treadmilling spherical populations \textit{fixed} at the initial radii $R_0$ of the inflating expansions. Note that the inflating and treadmilling expansions have different scaling variables, shown along the vertical and horizontal axes with the inflationary case shown first. In (a), we see that contrary to an inflating expansion, mutations always lead to the extinction of a $m=0$ strain in a treadmilling population at the  transition. In (b), we find that the $m=0$ cluster saturates at a finite size related to the population front size in a treadmilling tumor. At early times, the treadmilling and inflating tumors have similar scaling properties.  }
\end{figure}

To highlight the differences, we now compare the scaling functions for inflating spherical expansions to those for a \textit{treadmilling} population on a sphere (see figure~\ref{fig:treadmillsingleseed}).  In a treadmilling population, the finite size of the population introduces strong corrections to the scaling behavior at long times. However, the corrections are quite different from the corrections due to inflation.  For example, at long times, the one-way mutations lead to an even more rapid (exponential in time) extinction of the fittest $m=0$ cells at the transition, as shown in figure~\ref{fig:treadmillsingleseed}(a).  By contrast, inflation is able to \emph{save} the $m=0$ sector with a non-zero probability. Also, we see in figure~\ref{fig:treadmillsingleseed}(b) that the sector size saturates due to the finite population front size in a treadmilling population.  In the inflating case, the sector grows even more rapidly at times $t \gg t^*$.

\section{\label{sec:conclusions}Conclusions}

Range expansions near a fitness collapse transition present a fascinating example of a non-equilibrium critical phenomenon.   We have shown how two- and three-dimensional range expansions with thin, actively growing frontiers exhibit multicritical scaling behavior which may be characterized by coupled directed percolation processes.   We proposed a stochastic partial differential equation hierarchy to describe these expansions and analyzed the hierarchy using mean-field theory.  For three-dimensional expansions, we varied the irreversible, deleterious mutation rate $\mu$ and the strength $s$ of the mutation to calculate the shape of the phase diagram. We pointed out key differences from two-dimensional spatial range expansions and examined the effect of varying the possible maximum number of accumulated mutations, $M-1$ ($M$ total species).    In particular, the weaker effects of genetic drift  yield a more stable fitness distribution in three dimensions, with fitness collapse occurring in a smaller region of the phase space.   We also found that increasing $M$  leads to possible deviations (or a slow crossover) from the expected directed percolation scaling predictions.

We  also considered how inflating population frontiers modify the fitness collapse transition for the case of just a single deleterious mutation ($M=2$ total species), which is relevant for studying slightly deleterious passenger mutations in cancerous tissue.  We  find dramatic differences between treadmilling and inflating expansions, similar to those highlighted in \cite{MLSphere}.   We expect that the broad features of our results, such as the enhanced survival probability and cluster size scaling in the inflationary regime, will survive in a more realistic model with many accumulating mutations ($M \rightarrow \infty$).

 There is much room for future work.  For example, simulations of larger range expansions would be necessary to confirm that we are not seeing transient behavior and a universality class genuinely different from directed percolation in three-dimensional expansions.  An exploration of the proposed equation for the evolutionary dynamics, equation~(\ref{eq:UCDPLang}), (via a renormalization group analysis or a similar technique) would also be helpful in this context.
Another interesting direction would be to introduce a more realistic growth dynamics that allows the range expansion to develop a rough front. These rough fronts can couple strongly to the evolutionary dynamics and lead to profound changes in the behavior of the evolution near non-equilibrium phase transitions \cite{frey, MLDRNPRL}.

\section*{Acknowledgements}

The author is deeply grateful to D. R. Nelson and U. C. T\"auber for a critical reading of the manuscript and helpful discussions. This work was supported in part by the National Science Foundation (NSF) through grant DMR-1306367, and by the Harvard Materials Research Science and Engineering Center (DMR-1420570).  Support was also provided from NSF\ grant DMR-1262047. Computational resources were provided by the Harvard University Research Computing Group through the Odyssey cluster.

 \section*{References}

\bibliographystyle{iopart-num}
\bibliography{FitnessCollapse}

\providecommand{\newblock}{}
\begin{thebibliography}{10}
\expandafter\ifx\csname url\endcsname\relax
  \def\url#1{{\tt #1}}\fi
\expandafter\ifx\csname urlprefix\endcsname\relax\def\urlprefix{URL }\fi
\providecommand{\eprint}[2][]{\url{#2}}

\bibitem{lenski}
Elena S~F and Lenski R~E 2003 {\em Nat. Rev. Genet.\/} {\bf 4} 457--469

\bibitem{Ewens}
Ewens W~J 2004 {\em Mathematical Population Genetics\/} 2nd ed vol~I (New York:
  Springer)

\bibitem{KorolevRMP}
Korolev K~S, Avlund M, Hallatschek O and Nelson D~R 2010 {\em Rev. Mod.
  Phys.\/} {\bf 82} 1691--1718

\bibitem{cancermodel}
Araujo R~P and McElwain D~L~S 2004 {\em B. Math. Biol.\/} {\bf 66}(5)
  1039--1091

\bibitem{KorolevBac}
Korolev K~S, Xavier J~B, Nelson D~R and Foster K~R 2011 {\em Am. Nat.\/} {\bf
  178} 538--552

\bibitem{KorolevMueller}
Korolev K~S, M\"uller M~J~I, Karahan N, Murray A~W, Hallatschek O and Nelson
  D~R 2012 {\em Phys. Biol.\/} {\bf 9} 026008

\bibitem{animalRE}
Thomas C~D, Gillingham P~K, Bradbury R~B, Roy D~B, Anderson B~J, Baxter J~M,
  Bourn N~A~D, Crick H~Q~P, Findon R~A, Fox R, Hodgson J~A, Holt A~R, Morecroft
  M~D, O'Hanlon N~J, Oliver T~H, Pearce-Higgins J~W, Procter D~A, Thomas J~A,
  Walker K~J, Walmsley C~A, Wilson R~J and Hill J~K 2012 {\em PNAS\/} {\bf 109}
  14063--14068

\bibitem{spatialratchet}
Otwinowski J and Krug J 2014 {\em Phys. Biol.\/} {\bf 11} 056003

\bibitem{MKNPRE}
Lavrentovich M~O, Korolev K~S and Nelson D~R 2013 {\em Phys. Rev. E\/} {\bf 87}
  012103

\bibitem{nelsonhallatschek}
Hallatschek O and Nelson D~R 2010 {\em Evolution\/} {\bf 64}(1) 193--206

\bibitem{ratchet1}
Muller H~J 1964 {\em Mutat. Res.-Fund. Mol. M.\/} {\bf 1}(1) 2--9

\bibitem{ratchet2}
Haigh J 1978 {\em Theor. Popul. Biol.\/} {\bf 14}(2) 251--267

\bibitem{hochberg}
Folkman J and Hochberg M 1973 {\em J. Exp. Med.\/} {\bf 138} 745--753

\bibitem{NEQPTBook}
Henkel M, Hinrichsen H and L\"ubeck S 2008 {\em Non-Equilibrium Phase
  Transitions\/} vol I - Absorbing Phase Transitions (The Netherlands: Springer
  Science)

\bibitem{Hinrichsen}
Hinrichsen H 2000 {\em Adv. in Phys.\/} {\bf 49}(7) 815--958

\bibitem{KorolevCancer}
McFarland C~D, Korolev K~S, Kryukov G~V, Sunyaev S~R and Mirny L~A 2013 {\em
  PNAS\/} {\bf 110} 2910--2915

\bibitem{viralmut}
Domingo E and Holland J~J 1997 {\em Annu. Rev. Microbiol.\/} {\bf 51} 151--178

\bibitem{brainbow1}
Paddock S 2008 {\em Biotechniques\/} {\bf 44} 643--647

\bibitem{brainbow2}
Hawley T~S, Telford W~G and Hawley R~G 2001 {\em Stem Cells\/} {\bf 19}(2)
  118--124

\bibitem{brainbow3}
Buckingham M~E and Meilhac S~M 2011 {\em Dev. Cell\/} {\bf 21}(3) 394--409

\bibitem{eigen}
Eigen M, McCaskill J and Schuster P 1988 {\em J. Phys. Chem.\/} {\bf 92}
  6881--6891

\bibitem{MLSphere}
Lavrentovich M~O and Nelson D~R 2015 {\em Theor. Popul. Biol.\/}
  10.1016/j.tpb.2015.03.002

\bibitem{generallangevin}
Good B~H and Desai M~M 2013 {\em Theor. Popul. Biol.\/} {\bf 85} 86--102

\bibitem{brenneryeast}
Nguyen B, Upadhyaya A, van Oudenaarden A and Brenner M~P 2004 {\em Biophys.
  J.\/} {\bf 86}(5) 2740--2747

\bibitem{frey}
Kuhr J~T, Leisner M and Frey E 2011 {\em New J. Phys.\/} {\bf 13} 113013

\bibitem{MLDRNPRL}
Lavrentovich M~O and Nelson D~R 2014 {\em Phys. Rev. Lett.\/} {\bf 112} 138102

\bibitem{Bennett}
Bennett C~H 1972 {\em J. App. Phys.\/} {\bf 43} 2727--2734

\bibitem{domany}
Domany E and Kinzel W 1984 {\em Phys. Rev. Lett.\/} {\bf 53} 311--314

\bibitem{gardiner}
Gardiner C~W 1985 {\em Handbook of Stochastic Methods\/} 2nd ed (Berlin:
  Springer-Verlag)

\bibitem{kimurapaper}
Kimura M and Weiss G 1964 {\em Genetics\/} {\bf 49}(4) 561--576

\bibitem{RougheningCoupledDP}
Alon U, Evans M~R, Hinrichsen H and Mukamel D 1998 {\em Phys. Rev. E\/} {\bf
  57} 4997--5012

\bibitem{JanssenCoupledDP}
Janssen H~K 2001 {\em J. Stat. Phys.\/} {\bf 103}(5-6) 801--839

\bibitem{TauberCoupledDP}
T\"auber U~C, Howard M~J and Hinrichsen H 1998 {\em Phys. Rev. Lett.\/} {\bf
  80} 2165--2168

\bibitem{TauberCoupledDP2}
Goldschmidt Y~Y, Hinrichsen H, Howard M and T\"auber U~C 1999 {\em Phys. Rev.
  E\/} {\bf 59} 6381--6408

\bibitem{AsymmCoupledDP}
Noh J~D and Park H 2005 {\em Phys. Rev. Lett.\/} {\bf 94} 145702

\bibitem{threshold1}
Biebricher C~K and Eigen M 2005 {\em Virus Res.\/} {\bf 107}(2) 117--127

\bibitem{functionology}
Wilf H~S 1994 {\em generatingfunctionology\/} 2nd ed (San Diego: Academic
  Press)

\bibitem{gerlandhwa}
Gerland U and Hwa T 2002 {\em J. Mol. Evol.\/} {\bf 55}(4) 386--400

\bibitem{wagner}
Wagner G~P and Krall P 1993 {\em J. Math. Bio.\/} {\bf 32}(1) 33--44

\bibitem{voter1}
Liggett T~M 1985 {\em Interacting Particle Systems\/} (New York:
  Springer-Verlag)

\bibitem{voter2}
Dornic I, Chat\'e H, Chave J and Hinrichsen H 2001 {\em Phys. Rev. Lett.\/}
  {\bf 87} 045701

\bibitem{CDPFT}
Janssen H~K 2005 {\em J. Phys.: Condens. Matter\/} {\bf 17} S1973--S1993

\bibitem{treadmill1}
Cheng G, Tse J, Jain R~K and Munn L~L 2009 {\em PLoS ONE\/} {\bf 4}(2) e4632

\end{thebibliography}

\end{document}